\documentclass{elsart}
\usepackage{graphics}
\usepackage{amsmath}
\usepackage{amssymb}
\usepackage{epsfig}

\begin{document}
\begin{frontmatter}
\title{Localization of a pair of bound particles in a
       random potential}

\author{M. Turek\thanksref{MT_email}},
\thanks[MT_email]{Present address: University of Regensburg,
 Institute for Theoretical Physics, D-93040 Regensburg, GERMANY}
\ead{marko.turek@physik.uni-regensburg.de}
\author{W. John}
\ead{john@theory.phy.tu-dresden.de}

\address{Institut f{\"u}r Theoretische Physik,
           Technische Universit{\"a}t Dresden,
           D-01062, Germany}
\date{\today}

\begin{abstract}
We study the localization length $l_c$ of a
pair of two attractively bound 
particles moving in a one-dimensional random potential. We show in which
way it depends on the interaction potential between the
constituents of this composite particle. For a pair with many 
bound states $N$ the localization length 
is proportional to $N$, independently
of the form of the two particle interaction.
For the case of two bound states we present
an exact solution for the corresponding
Fokker-Planck equation and demonstrate
that $l_c$ depends sensitively on  the shape of the interaction
potential and the symmetry of the
bound state wave functions.
\end{abstract}

\end{frontmatter}

%
\section{Introduction}
\label{Introduction}
Within the single-parameter scaling hypothesis \cite{Abrahams79}
the dimensionless conductance $g(L)$ as a function of system size
$L$ determines the transport properties of noninteracting
electrons in a random potential (for a review, see 
Kramer and MacKinnon \cite{Kramer93}).
In one-dimensional
disordered systems all states are localized \cite{Mott61} and the
conductance $g(L)$ decreases exponentially \cite{Anderson80} with increasing
sample size $L$. The localization length $l_c$ can
be derived from the asymptotic behavior of the conductance
$g(L)$ using the relation
\begin{equation}
l_c^{-1} = - \lim_{L \to \infty} \frac{d}{dL} \ln g(L).
\end{equation}
The localization length $l_c$ and the Ljapunov exponent
$\gamma \sim l_c^{-1}$ are self-averaging, i.e.
non-random quantities \cite{Lifshitz88} where\-as the distribution
function of the conductance is approximately log normal
\cite{Abrikosov81}. In the weak disorder limit characterized
by a small Fermi wavelength compared to the scattering length
$l$ the localization length
$l_c$ is given by the mean free path $l$ for backward
scattering \cite{Lifshitz88}.

In this paper we study the localization of a composite
particle with internal degrees of freedom. As a model
we consider a single pair of two particles in a
weak one-dimensional white noise potential. We restrict
our considerations to the limit of two tightly bound
constituents so that the
pair does not decay during its motion through
the disorder potential. Considering the
semiclassical limit the kinetic energy of
the center of mass motion is assumed to be larger than the typical
energy of the disorder potential. 
Typical length scales of our model
include the mean free paths $l_b$ and $l_f$ of the pair for
backward and forward scattering, respectively. The scattering
processes cause a transfer of kinetic energy of the center
of mass motion to the relative motion within the pair and
vice versa. Therefore, the scattering lengths $l_b$ and
$l_f$ depend on the structure of the pair, i.e. on the
interaction potential between the constituents.
Thus one expects that the pair localization
length $l_c$ also depends on the shape of
the attractive pair potential.

As another but different example for the interplay between
interaction and disorder
many recent papers study the coherent propagation of two
interacting particles (TIP) in a one-dimensional random
potential (see \cite{TIP}
and the literature cited
there). It is worth mentioning that the TIP model considers
the propagation of two repulsing or weakly attracting
particles which form a pair due to the scattering by the
disordered potential. The size of the pair
is of the order of the one particle localization length
$l$. As Shepelyansky \cite{Shepelyansky94} and Imry
\cite{Imry95} argued, attractive as well as repulsive
interaction between the particles cause a localization length
of the pair which is much larger than the single particle
localization length. In the present work, we study
two particles with strong interaction which form bound
states and have a size that is much smaller than the
one particle localization length. For that reason the TIP model
and the composite particle model studied in this work
consider different limiting cases, which are not directly comparable.
Moreover, in the mentioned TIP approach a
lattice model of two particles interacting
at one lattice site is considered. Generally the coherent motion
of the interacting pair is studied only in the middle of
the band ($E=0$). However, we study a continuum model of two
interacting particles in the semiclassical regime with a
Fermi wavelength much smaller than the scattering length
($\lambda_F \ll l$).


A composite particle with $N$ bound states may be considered as an
$N$-channel problem for the center of mass motion. On the other
hand the motion of a single electron in a thick wire is also a
multichannel problem since the transversal motion is quantized.
In the thick wire the scattering processes between different channels
are random and of the same strength. The localization length 
$l_c$ of the wire increases with the channel number $N$ as
$l_c \sim N l$ because
effectively only one of these channels gives rise to coherent
backward scattering and produces 
localization \cite{Dorokhov83,Beenakker97}. In our
composite particle model the scattering processes between different
bound states are also random. However, in contrast to the
model of a thick wire, the scattering probabilities depend
strongly on the channel numbers. Therefore it is interesting to
investigate the localization length in the
composite particle model and compare the results for
$l_c$ of both different models.

The model of two particles with strong attractive
interaction might be relevant for the Coulomb-correlated electron-hole
pairs in disordered semiconductors. Recent work \cite{Brinkmann99} has
dealt with the dynamics  of the electron-hole pair in a 1D model.
Another physical problem which can be related to the model we consider
below is the superconductor-isolator transition. In order to study
the effects of interaction and disorder, Lages and 
Shepelyansky \cite{Lages} investigated 
numerically the Cooper problem of two quasiparticles with
attractive interaction above a frozen Fermi sea in the presence of
disorder.

The basic ideas of the model we consider 
were introduced by Dorokhov \cite{Dorokhov90}.
He studied a pair of particles bound by a harmonic oscillator
potential in a weak random disorder
potential. For this very special interaction he found
a dominant forward scattering in the case of many relevant
bound states $N$ ($l_b/l_f \sim \ln N$) employing the specific
properties of the harmonic oscillator eigenfunctions. Moreover,
the pair scattering length $l_b$ was shown to be 
on the order of the single particle mean free path $l$ and
independent of the strength of the oscillator potential.
Using the fact that the forward scattering is 
the dominant process the pair localization length
$l_c = N l_b / 2 \sim N l$ was calculated.

Our results show that Dorokhov's conclusions
are generic for any attractive interaction potential
in the case of many bound states $N \gg 1$ regardless
of its shape.
We extend the method \cite{Dorokhov90} from
the harmonic oscillator potential to an arbitrary attractive
interaction potential. This can be accomplished
since the excited states of the
relative motion within the pair can be described in the
semiclassical approximation. We show that the specific
properties of the harmonic oscillator wave functions
(in \cite{Dorokhov90} the result was obtained using
the recurrence relations of the Hermitian polynomials)
are of no importance in the
multi-channel case. Using WKB wave functions for
the bound states of the pair we estimate the scattering
matrix elements and the mean free paths $l_b$ and $l_f$
in the case $N \gg 1$. We find that $l_b \gg l_f$
for any interaction potential. We also show that
$l_b$ is qualitatively independent of the interaction.
For the localization length $l_c$ we obtain $l_c \sim N l_b / 2$
which implies that $l_c$ is also independent of the interaction.

In addition to the considerations for
many bound states $N \gg 1$ we also investigate the first 
nontrivial case of a small number of bound states, namely
$N=2$. For a single particle in
a random white noise potential the mean
free paths $l_b$ and $l_f$ are equal.
We show that already for a pair with only two bound
states, the forward scattering is enhanced relative to the
backward scattering and that the ratio $l_f/l_b$ depends
sensitively on the energy.
Finally we derive an exact expression for the
localization length of the pair with two accessible bound states.
In contrast to earlier treatments \cite{Dorokhov83}
of this problem
we do not have to impose any additional restrictions to the 
scattering matrix elements. Therefore, our solution allows
to study the influcence of the two particle interaction
on the localization directly. The result indicates
that no simple relation exists between $l_c$ and the mean free
path of the pair $l_b$ as it was in
the limit of many channels $N \gg 1$.
As an application of our
general result we present numerical calculations
showing that the localization length $l_c$
depends sensitively on the interaction potential,
the symmetry of the bound state wave
functions and the total energy of the pair.

The paper is organized as follows. In Section
\ref{Model_and} we describe the model and briefly
summarize the transfer matrix approach 
applied to the composite particle model with
arbitrary interaction potential between the two
constituents. In Section
\ref{Multi-channel} we consider a pair with high enough
total energy allowing for many bound
states. We determine the scattering lengths $l_b$ and
$l_f$ as well as the pair localization length $l_c$.
In Section \ref{Two_channel}
we derive an exact expression for the localization length
of a pair with two bound states ($N=2$). Furthermore, we
present numerical results for the localization length $l_c$
as well as for the mean free paths $l_b$ and $l_f$ 
for different interaction potentials.

\section{Model and method of calculations}
\label{Model_and}
We consider two interacting particles in a one-di\-men\-sio\-nal
random potential $V(x)$. The Hamiltonian is given
by
\begin{equation}
\label{H2}
 {\mathcal H} = -\frac{1}{4} \frac{\partial ^2}{\partial x^2} -
   \frac{\partial ^2}{\partial y^2}
   + u(y) + V(x+y/2) 
   + V(x - y/2),
\end{equation}
where the units are chosen in a way that $\hbar=1$ and the
single electron mass $m=1$.
The variable $x$ denotes the center of mass
coordinate of the pair
while $y$ represents the relative coordinate.
The symmetric interaction potential $u(y)$ between the two particles
is assumed to be attractive. The bound states $\phi_n(y)$ and
the corresponding energy eigenvalues $\epsilon_n$ are given
by
\begin{equation}
\label{SGL:relat}
 \left[ - \frac{\partial^2}{\partial y^2} 
 + u(y) \right] \phi_n(y) = \epsilon_n \phi_n(y).
\end{equation}
In the case of two identical particles the states $\phi_n(y)$
of even parity describe two bosons or a singlet state of fermions
while the wave functions of odd parity correspond to a triplet state
of fermions.
The random potential $V(x)$ is nonzero in the finite interval
$0<x<L$. This region of length $L$ is enclosed by two ideal leads
without any disorder.

The scattering of the center of mass of the
pair with $N$ bound states can be mapped onto an $N$-channel
problem of a single free particle. The stationary scattering
states of the pair in the leads are given by
\begin{equation} 
\label{Ansatz2} 
  \Psi(x,y)=\sum\limits_{n=0}^{N-1} \phi_n(y) \left[ A_n
  \frac{\exp{(i k_n x)}}{\sqrt{k_n}} + 
  B_n \frac{\exp{(- i k_n x)}}{\sqrt{k_n}} 
  \right].
\end{equation}
Here, the momentum of the center of mass is denoted
by $k_n$. The total energy of the pair
$E = k_n^2/4 + \epsilon_n$ determines the number
of bound states (channels) $N$ which are included
in the sum in Eq. (\ref{Ansatz2}). For a given
total energy $E$ this number $N$ can be obtained from
the relation $\epsilon_{N-1} < E < \epsilon_N$.

As far as the coefficients $A_n$ and $B_n$ are concerned
one has to distinguish between the left lead and the right
lead. They can be written
as vectors $(\vec{A}^L,\vec{B}^L)$ and
$(\vec{A}^R,\vec{B}^R)$ for the left and right one, respectively.
The transfer matrix
$\tau$ describes the scattering in the region between the two leads.
It relates the states in the right lead to the
states in the left lead \cite{Beenakker97}:
\begin{equation}
\label{Def:tau}
\left( \begin{array}{c}
        \vec{A}^L\\ \vec{B}^L
        \end{array}
\right) = \tau
\left( \begin{array}{c}
        \vec{A}^R\\ \vec{B}^R
        \end{array}
\right).
\end{equation}
The transfer matrix $\tau$ is a $2N \times 2N$ symplectic
matrix which can be expressed by the $N \times N$ reflection
matrix $r$ and the transmission matrix $t$. These
matrices describe the scattering between the channels
$n$ and $m$ which correspond to the associated bound states
of the pair. The dimensionless conductance $g$ can be 
expressed \cite{Imry97} in terms of the $N$ eigenvalues
$T_n$ of the Hermitian matrix $t^\dagger t$ as $g=\sum T_n$.
Therefore, it is convenient to
employ the Hermitian matrix $M\equiv \tau^\dagger \tau$
which can be expressed by the polar decomposition as
\cite{Dorokhov83}
\begin{equation}
\label{param.:M}
M = \left( \begin{array}{cc} u & 0 \\ 0 & u^*
              \end{array} \right)
       \left( \begin{array}{cc} \cosh 2\Gamma & 
              - \sinh 2\Gamma \\ - \sinh 2\Gamma & \cosh 2\Gamma
              \end{array} \right)
       \left( \begin{array}{cc} u^\dagger & 0 \\ 0 & u^T
              \end{array} \right).
\end{equation}
In this parameterization $u$ is
a $N \times N$ unitary matrix and $\Gamma$ is
a $N \times N$ diagonal matrix. The transmission
eigenvalues $T_n$ are then related to the
matrix elements $\Gamma_n$ by
$T_n = 2/[\cosh (2 \Gamma_n) + 1]$.

To evaluate the transmission eigenvalues $T_n(L)$ for
a given size $L$ we now consider
the matrix $M$ as a function of the sample
size $L$. As was shown in \cite{Dorokhov90} the evolution
of $M(L)$ with increasing $L$ can be described by a set of
Langevin equations. The random potential $V(L)$ enters
these equations as a multiplicative factor. In order to
derive these equations the following
assumptions were made:
\begin{itemize}
\item The potential $V(x)$ is a weak Gaussian white noise
potential with the correlation function
\begin{equation}
\label{correlation:V}
\langle V(x) V(x') \rangle = 2D \delta (x-x'),
\end{equation}
where $\langle \dots \rangle$ denotes the ensemble
average. The mean free path $l$ of the pair in comparison to the
inverse wave numbers $k_n^{-1}$ is the appropriate 
quantity to describe the strength of the disorder. The 
semiclassical limit is considered where $k_n l \gg 1$.
The mean free path $l$ of a single particle with Fermi
energy $E_F$ is given by $l=E_F / D$.
\item The mean free path $l$ is large compared to the
typical width $b_n$ of the bound states $\phi_n(y)$:
$l \gg b_n$. This assumption ensures the concept of
mean free path for the center of mass motion.
\end{itemize}

The localization length $l_c$ of the pair is studied
by the asymptotic behavior
of the transmission eigenvalues $T_n(L)$ for large
system sizes $L \to \infty$. These
coefficients $T_n(L)$ decrease exponentially
with the system size $L$. 
The corresponding length scales are defined by
\begin{equation}
\label{Def:l_n}
  l_n^{-1} \equiv - \lim\limits_{L \to \infty} \frac{d}{dL}
  \langle \ln T_n \rangle.
\end{equation}
The transmission eigenvalues can be arranged 
in a hierarchical order
$T_{N-1} \ll T_{N-2} \ll \dots \ll T_0 \ll 1$.
The localization length $l_c$ is then given by the largest
length $l_0$. 
All lengths $l_n$ averaged over the $N$ channels just
give the mean free path $l_b$ of the center of mass
for backward scattering \cite{Dorokhov83}:
\begin{equation}
\label{avg.l_n}
 \frac{1}{N} \sum\limits_{n=0}^{N-1} l_n^{-1} = l_b^{-1}
\end{equation}
(see Eq. (\ref{l_b,f}) below
for the backward scattering length).
Taking into account the exponential smallness of all
transmission eigenvalues and their different order of
magnitude, the Langevin equations for the matrix
elements $\Gamma_n$ and $u_{nm}$ are given by
\begin{eqnarray}
\label{LG:gamma_n}
 \frac{\partial \Gamma_n}{\partial L} &=& i
 \left( \beta_u^* - \beta_u \right)_{nn} \cdot V(L) \\
\label{LG:u_nm}
 \frac{\partial u_{nm}}{\partial L} &=& -2 i 
 \sum\limits_{l=0}^{N-1}
 \left[ \alpha_{nl} u_{lm} + u_{nl}(\beta_u^*)_{lm} 
 \Theta(l-m) \right. \nonumber \\
 & & \quad \quad \quad \quad
 \left. + \; u_{nl}(\beta_u)_{lm} \Theta(m-l) \right]
 \cdot V(L)
\end{eqnarray}
where the abbreviation $\beta_u$ stands for
$\beta_u \equiv u^\dagger \beta u^*$. The value
of the step function $\Theta(k)$ for $k=0$ is defined
as $\Theta(0) \equiv 1/2$. The matrix elements $\alpha_{nm}$
and $\beta_{nm}$ in Eq. (\ref{LG:u_nm})
describe the forward scattering and the backward
scattering of the pair, respectively. They depend on the
two particle interaction and can be
written as
\begin{eqnarray}
\label{W_nm:alpha,beta}
 \alpha_{nm} &=& \frac{e^{i(k_n-k_m) L}}{\sqrt{k_n k_m}}
  W_{nm}(k_n - k_m) , \nonumber \\
 \beta_{nm}  &=& \frac{e^{-i(k_n+k_m) L}}{\sqrt{k_n k_m}}
  W_{nm}(k_n + k_m) ,
\end{eqnarray}
where $W_{nm}(k)$ is
\begin{equation}
\label{W_nm}
 W_{nm}(k) \equiv 2 \int\limits_{-\infty}^{\infty}
 dy \; \phi_n(y) \phi_m(y) \, \cos \left( \frac{ky}{2} \right).
\end{equation}
Within the Born approximation, the mean free
paths $l_{b,f}$ of the pair for backward and
forward scattering are given by \cite{Dorokhov90}:
\begin{equation}
\label{l_b,f}
 l_b^{-1} = \frac{8 D}{N} \sum\limits_{n,m}^{N-1} 
  | \beta_{nm} |^2 \quad \text{and} \quad
 l_f^{-1} = \frac{8 D}{N} \sum\limits_{n,m}^{N-1}
  | \alpha_{nm} |^2.
\end{equation}
Note that one has to consider bound
states of a given parity only. Due to the finite width
of the pair we therefore expect that the
forward scattering is enhanced relative to the backward
scattering although the scattering of a single particle
is isotropic.
This circumstance allows to solve the many channel
problem \cite{Dorokhov90} with $N\gg 1$ (see section \ref{Multi-channel}).

Using the fact that the transport is effectively
determined by just one channel ($T_0 \gg T_{n>0}$)
Eq. (\ref{Def:l_n}) results in the following
expression for the pair localization
length \cite{Dorokhov90}
\begin{equation}
\label{l_c:general}
 l_c^{-1} = 16D \sum\limits_{n,m=0}^{N-1} 
 \langle |u_{n0}|^2 |u_{m0}|^2 \rangle (1-\frac{1}{2}\delta_{nm})
 |\beta_{nm}|^2.
\end{equation}
Hence, the localization length $l_c$ depends on the correlation
function $\langle |u_{n0}|^2 |u_{m0}|^2 \rangle$ which
has to be evaluated using the nonlinear Langevin Eqs. (\ref{LG:u_nm}).

For $N=1$ the unitary matrix $u$ is simply a phase factor
with $|u_{00}|=1$. The localization length $l_c$ is then given
by the mean free path $l_b$ for backward scattering
\begin{equation}
\label{l_c:N=1}
l_c^{-1} = l_b^{-1} = 8D |\beta_{00}|^2
\end{equation}
and the interaction potential $u(y)$ enters via
the ground state wave function in the matrixelement
(\ref{W_nm}). In the next section we consider the
multi-channel case $N \gg 1$ and show that the
localization length depends only weakly on the
interaction.

\section{Multi-channel localization length}
\label{Multi-channel}
In order to evaluate the localization length (\ref{l_c:general})
in the multi-channel case $N \gg 1$ one has to compute
the correlation function $\langle |u_{n0}|^2 |u_{m0}|^2 \rangle$
using the nonlinear Langevin equations (\ref{LG:u_nm}). 
For the harmonic oscillator interaction it was shown
in \cite{Dorokhov90} that the
forward scattering dominates in Eq. (\ref{LG:u_nm}).
Therefore the nonlinear terms due to the backward scattering
processes were neglected in (\ref{LG:u_nm}) and it was
found that the correlation function
is properly described by the invariant
unitary ensemble
\begin{equation}
\label{correlation:u}
\langle |u_{n0}|^2 |u_{m0}|^2 \rangle = (1 + \delta_{nm})/N^2.
\end{equation}
In the following we proof that the same result is valid
for any attractive interaction potential.
The order of magnitude of forward and backward scattering
processes is determined by the corresponding mean free
paths $l_f$ and $l_b$ given by (\ref{l_b,f}).
These mean free paths depend on the pair interaction
potential $u(y)$ via the bound state wave functions
$\phi_n(y)$ in Eq. (\ref{W_nm}). We calculate the matrix elements
$W_{nm}(k)$ using WKB wave functions and
evaluate the integral (\ref{W_nm}) in the saddle
point approximation.
This approximation is justified by the semiclassical
approximation and the assumption $N \gg 1$, which implies that
the vast majority of bound states is properly described by the
WKB approximation.
In this approximation the following expression
can be derived from (\ref{W_nm})
\begin{eqnarray}
\label{WKB:W_nm}
 W_{nm}(k) = &4& (-1)^{(n \pm m)} \left(
 \frac{2 \pi}{{\mathcal T}_n {\mathcal T}_m |u'(y_0) \cdot k|} \right)^{1/2}
 \nonumber \\
 &\times & \cos \left[ \mp \frac{1}{2} \int\limits_{k_{2,1}}^{|k|}
 d\tilde{k} \; y_0(\tilde{k}) - \frac{\pi}{4} \right].
\end{eqnarray}
The time ${\mathcal T}_n$ denotes the period of the semiclassical motion
of the bound state $n$. The two saddle points $\pm y_0$ are functions
of $k$ and follow from the implicite equation
\begin{equation}
 \label{WKB:u(y_0)}
 u(y_0) =
  \frac{\epsilon_n+\epsilon_m}{2} - \frac{k^2}{16}
  -\frac{(\epsilon_n - \epsilon_m)^2}{k^2}.
\end{equation}
In Eq. (\ref{WKB:W_nm}) $u'(y_0)$ stands for the derivative
of the potential evaluated at position $y_0$. It results
from the Gaussian integral at the saddle point. 
Calculating the matrix elements
$W_{nm}(k)$ for backward or forward scattering we
set $k=k_n \pm k_m$. In this relation as well as in
Eq. (\ref{WKB:W_nm}) the upper sign and $k_2$ characterize
the backward scattering while the lower sign and $k_1$
represent the forward scattering. The expression
(\ref{WKB:W_nm}) is valid within the interval $k_1 < |k| < k_2$
where Eq. (\ref{WKB:u(y_0)}) has a real solution $y_0(k)$.
The limits of the interval are given by
\begin{equation}
\label{k_1,2}
 k_{2,1} = 2 (\sqrt{\epsilon_n-u_0} \pm \sqrt{\epsilon_m-u_0})
\end{equation}
with the potential minimum $u_0 \equiv u(y=0)$. Outside this
$k$ region the saddle point is shifted to the complex
$y$-plane and the matrix element $W_{nm}(k)$ decreases
exponentially \cite{LLBd3}. Calculating the mean free paths
(\ref{l_b,f}) we will neglect these exponentially small
contributions.

\begin{figure}
\centerline{
\epsfxsize=0.47\textwidth
\epsfbox{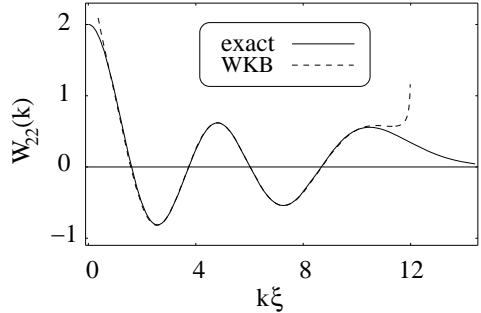}}
\caption{Comparison of the exact function $W_{22}(k)$ and
 the calculated WKB estimate in the case of the harmonic
 oscillator interaction. The WKB approximation is
 excellent already for small quantum numbers in the
 wave function (in this example $\phi^s_{n=2}$) as long
 as $k$ lies in the oscillating regime. The WKB approximation
 clearly breaks down at $k_1 \xi =0$ and
 $k_2 \xi = 12$.}
\label{fig:w_22}
\end{figure}

To demonstrate the accuracy of the WKB wave functions
together with the saddle point approximation
we show an example for a matrix element
in Fig. \ref{fig:w_22}. We have calculated
the exact function $W_{22}(k)$ for the harmonic oscillator
potential $u_{osc}(y) = \xi^{-2} (y/\xi)^2$ using
the analytical expression for the third symmetric
bound state $\phi^s_2(y)$. This exact
result is compared with our WKB result (\ref{WKB:W_nm}).
One clearly sees that the oscillatory part
is very well described by the approximation. 
Of course, close to $k=k_{1,2}$ the saddle point approximation
fails because $u'(y_0)$ goes to zero and the Gaussian
approximation is not valid. Furthermore, the diagonal matrix
elements $W_{nn}$ cannot be calculated from Eq. (\ref{WKB:W_nm})
for $k=0$. However, definition (\ref{W_nm}) immediately
yields $W_{nn}(k=0) = 2$.

%
%

The scattering lengths are then evaluated
by replacing the sum in (\ref{l_b,f}) by an integral (with
$k_n \to k$ and $k_m \to \tilde{k}$).
We obtain
\begin{equation}
\label{l_b,f:integral}
l_{b,f}^{-1} = \frac{8D}{4 \pi N} \int
 \frac{dk \; d\tilde{k}}{(k \pm \tilde{k}) \cdot |u'(y_0)|}.
\end{equation}
This result may be understood as follows. For $k^2=(k_n \pm k_m)^2$
the saddle point equation (\ref{WKB:u(y_0)}) can
be rewritten as
\begin{equation}
\label{saddlepoint}
u(y_0) = E - \frac{k_n^2}{4} - \frac{k_m^2}{4} = \epsilon_n
 - \frac{k_m^2}{4} = \epsilon_m - \frac{k_n^2}{4}.
\end{equation}
This means that the kinetic energy of the relative
motion $\epsilon_n-u(y_0)$ for the initial channel $n$
transforms into the kinetic energy $k_m^2/4$ of the
center of mass in the final channel $m$ and vice versa.
The transition rate into the final channel $m$ is therefore
proportional to the probability density $\rho_n(k_m)$
of finding the pair in the initial channel $n$ with the
appropriate relative momentum. The expression (\ref{l_b,f:integral})
for the mean free paths follows then immediately
from $\rho_n(k_m)=4/({\mathcal T}_n |u'(y_0)|)$.

We will now show that Eq. (\ref{l_b,f:integral})
implies a mean free path for the backward scattering
that is almost
independent of the channel number $N$.
Using Eq. (\ref{saddlepoint})
we write $u'(y_0)$ formally as
\begin{equation}
\label{u'(y_0)}
u'(y_0) = F(u(y_0)) = F\left( 
 E - \frac{k^2}{4} - \frac{\tilde{k}^2}{4} \right).
\end{equation}
This is always a unique representation if 
the interaction potential is assumed to be monotonic.
Introducing polar coordinates the angular integration 
in (\ref{l_b,f:integral}) can be performed and we obtain
the expression for the backward scattering length
\begin{equation}
\label{WKB:l_b,result}
 l_b^{-1} = \sqrt{2} 
  \ln \left[ \frac{\sqrt{2}+1}{\sqrt{2}-1} \right]
 \frac{\int\limits_0^1 dt \,
           (1-t)^{-1/2} \, \left[ F(Et) \right]^{-1}}
 {\int\limits_0^1 dt \,
  (1-t)^{1/2} \, \left[ F(Et) \right]^{-1}} l^{-1}.
\end{equation}
The denominator in (\ref{WKB:l_b,result}) results from the
number $N$ of channels with given parity below the energy $E$.
Within the WKB approximation the relation between the number
$N$ and the energy $E$ can be expressed by using expression
(\ref{u'(y_0)}) once more:
\begin{eqnarray}
\label{WKB:N}
N(E) &=& \frac{1}{2 \pi} \int dy \, \sqrt{E - u(y)} \nonumber \\
     &=& \frac{1}{\pi} E^{3/2}
 \int\limits_0^1 dt \; (1-t^2)^{1/2} \,
 \left[ F(Et) \right]^{-1}.
\end{eqnarray}
From Eq. (\ref{WKB:l_b,result}) follows that $l_b \sim l$ with
a prefactor of order one. This prefactor depends only weakly
on the interaction potential $u(y)$ and the number of 
accessible bound states $N$. To demonstrate this
we consider two examples of interaction potentials.

In the case of a scale invariant potential
with $u(y)\sim y^\nu$ and $F(u) \sim u^{(\nu - 1)/ \nu}$ one immediately
realizes that $l_b / l$ is completely independent
of the energy as it was the case 
for the harmonic oscillator ($\nu=2$).
From (\ref{WKB:l_b,result}) follows
\begin{equation}
\label{u_si:l_b}
l_b^{-1} = l^{-1} 4 \sqrt{2} \ln \left[ 
 \frac{\sqrt{2}+1}{\sqrt{2}-1} \right] \cdot \left(
 \frac{1}{2} + \frac{1}{\nu} \right).
\end{equation}
This expression includes
the result for the harmonic oscillator
presented in
\cite{Dorokhov90} which was found by calculating
the matrix elements $W_{nm}$ exactly. For any
interaction potential $u(y)$ with many bound states
we expect only a weak dependence of $l_b$ on
the energy $E$ or on the number of channels $N$
which are below this pair
breaking energy. As a further example we give the result
for the P{\"o}schel-Teller interaction \cite{Junker96}
$u^a_{PT}(y)=\xi^{-2} (a^2-a(a+1)/\cosh^{2}[y/\xi])$
where $a$ is an integer parameter.
In contrast to the scale invariant potentials
this interaction has a maximal number $N_{max}$
of bound states given by the parameter $a$. 
The pair breaking energy is $\sim \xi^{-2}a^2$.
The semiclassical limit can be achieved
for a large parameter $a$ implying $N_{max} \gg 1$.
The backward scattering length (\ref{WKB:l_b,result})
can then be written as
\begin{equation}
\label{u_PT:l_b}
l_b^{-1} = l^{-1} \sqrt{2} \ln \left[
 \frac{\sqrt{2}+1}{\sqrt{2}-1} \right]
 \frac{1}{1-(N/N_{max})^2}.
\end{equation}
If the total energy of the pair is much smaller than
the pair breaking energy then
$N \ll N_{max}$ follows and again the ratio
$l_b/l$ is only very weakly energy dependent.

Using the same approach for the forward scattering
length $l_f$ one realizes that the main contribution
in Eq. (\ref{l_b,f:integral}) is due to the scattering
between adjacent channels with $k \approx \tilde{k}$.
However, the saddle point approximation fails for
$k=\tilde{k}$ and the integral in (\ref{l_b,f:integral})
diverges. Therefore, one has to exclude a small strip
of order $|k-\tilde{k}| \sim 1/N$ in the integral
and calculate the in-channel scattering separately.
We find in accordance with the ealier result for
the harmonic oscillator interaction \cite{Dorokhov90}
that
\begin{equation}
\label{WKB:l_f,result}
l_f^{-1} \sim l^{-1} \; \ln N
\end{equation}
for $N\gg 1$. From this estimate (\ref{WKB:l_f,result})
follows $l_f \ll l_b$ for the multi-channel case. This means
that the forward scattering is dominant and
the relation (\ref{correlation:u})
can be applied for any interaction
potential $u(y)$. The Eqs.
(\ref{WKB:l_b,result}, \ref{WKB:l_f,result}) are thus
the first major result of this work.

Using the correlation function (\ref{correlation:u}) we
then obtain the localization length
\begin{equation}
\label{final_result:l_c}
l_c = \frac{N}{2} l_b
\end{equation}
from expression (\ref{l_c:general}). Thus we find a
delocalization proportional to the number of
open channels $N$ compared to the one particle
localization length which equals $l$.
This shows that the localization
effect is qualitatively independent of the shape
of the attractive interaction potential $u(y)$. The interaction only
gives rise to a weak energy dependence and a numerical
prefactor in the backward scattering length of the pair $l_b$
compared to the single particle scattering length $l$.

Our result (\ref{final_result:l_c}) has the same
structure as in the different model 
of a thick wire \cite{Beenakker97}. 
For that case the Dorokhov-Mel\-lo-Pereyra-Kumar (DMPK)
equation describes the evolution of the distribution function
for the transmission eigenvalues $T_n$ (or $\Gamma_n$)
with increasing length of the wire.
Compared to our model an important simplification there
is the isotropy assumption for the scattering between
the channels on average. Using this assumption
the Fokker-Planck equation for the transmission
eigenvalues $T_n$ decouples from the distribution
function of the matrix elements $u_{nm}$. At least in the
absence of time-reversal symmetry the DMPK equation could be
solved exactly \cite{Beenakker94}. In our model the dominance
of the forward scattering processes is the important property 
which yields the correlation function (\ref{correlation:u}) and
leads with (\ref{l_c:general}) to the final result
(\ref{final_result:l_c}).

\section{Two channel system}
\label{Two_channel}
As yet another application of the
general result (\ref{l_c:general}) we now derive the
exact solution for the case of two open channels $N=2$.
Here, the matrix $u_{nm}$ is a $2 \times 2$ unitary matrix.
Thus only the expression
$\zeta \equiv |u_{00}|^2 - |u_{10}|^2$ occurs in
Eq. (\ref{l_c:general}) for the localization
length. One can rewrite
(\ref{l_c:general}) in terms of this variable
and obtains
\begin{equation}
\label{l_c:N=2,a}
 l_c^{-1} = 8D \left[ (\beta^+ - |\beta_{10}|^2)
   \langle \zeta^2 \rangle
  + 2 \beta^- \langle \zeta \rangle + 
  (\beta^+ + |\beta_{10}|^2) \right]
\end{equation}
with $\beta^\pm = (|\beta_{00}|^2 \pm |\beta_{11}|^2)/4$.
This reduces the ensemble average over combinations
of matrix elements $u_{nm}$
to the moments $\langle \zeta \rangle$
and $\langle \zeta^2 \rangle$. In order to find the
stationary distribution function $P(\zeta)$ we use
the parameterization $\zeta=\cos 2 \theta$ and
\begin{equation}
\label{param.:u}
u = \left( \begin{array}{cc} e^{-i \phi_0} & 0 \\
                             0 & e^{-i \phi_1}
           \end{array} \right)
    \left( \begin{array}{cc} \cos \theta & -\sin \theta \\
                             \sin \theta & \cos \theta
           \end{array} \right)
    \left( \begin{array}{cc} e^{-i \gamma} & 0 \\
                             0 & e^{i \gamma}
           \end{array} \right)
\end{equation}
for the unitary matrix $u_{nm}$. 
The stationary probability distribution can then be
obtained from the associated Fok\-ker-Planck equation.
Analyzing this equation we find that the exponentials
$e^{i \phi_{0,1}}$ and $e^{i \gamma}$ always
occur together with the matrix elements
$\alpha$ and $\beta$, respectively.
If one neglects the rapidly oscillating contributions
in this Fokker-Planck equation then only those terms which
contain the magnitudes $|\alpha_{nm}|^2$ and $|\beta_{nm}|^2$ remain.
For that reason, the stationary probability
distribution is independent of the phases
$\phi_{1,2}$ and $\gamma$. The solution $P(\zeta)$
of the stationary Fokker-Planck
equation follows from the first order differential
equation
\begin{eqnarray}
\label{DGL:P,N=2}
 0 &=& \left[ \frac{\partial}{\partial \zeta} 
  (1+A \zeta^2) + B \right] P(\zeta) \\
 \mbox{with} \quad
 A & \equiv& \frac{ |\beta_{10}|^2-\beta^+}{|\alpha_{10}|^2+\beta^+}
 \quad \text{and} \quad
 B \equiv \frac{2 \beta^-}{|\alpha_{10}|^2+\beta^+}.
\end{eqnarray}
It is of the type
$P(\zeta) \sim \exp [-B \int^\zeta dx (1+Ax^2)^{-1}]/(1+A \zeta^2)$.
Thus we arrive at the following expression for the
two channel localization length:
\begin{eqnarray}
\label{l_c:N=2}
 l_c^{-1} = 8 \, E \, l^{-1} ( |\alpha_{10}|^2 + \beta^+)
 & & \left[ 1 + 
 \frac{|\beta_{10}|^2+\beta^+}{|\alpha_{10}|^2+\beta^+}
 \right. \nonumber \\
 & & - \left. \frac{|\alpha_{10}|^2 + |\beta_{10}|^2}
      {|\alpha_{10}|^2+\beta^+} P^+ \right]
\end{eqnarray}
with $P^+ \equiv [P(1)+P(-1)]$. The length scale $l$
is the mean free path of a single particle having
the same energy as the pair, i.e. $E=2 E_F$.
The probabilities $P(\pm 1)$ for $\zeta = \pm 1$ can easily be obtained
from the solution of Eq. (\ref{DGL:P,N=2}).
It yields
\begin{eqnarray*}
 P^+ &=&
 \frac{B}{1+A} \coth \left[ \frac{B}{\sqrt{A}} \arctan \sqrt{A}
 \right] \quad \text{for} \quad A>0 \\
 P^+ &=& 
 \frac{B}{1+A} \coth \left[ \frac{B}{\sqrt{-A}} \mbox{artanh} \sqrt{-A}
 \right] \quad \text{for} \quad A<0.
\end{eqnarray*}
Expression (\ref{l_c:N=2}) provides a general solution
for the localization length in the two channel regime without
any further restrictions concerning the matrix elements
$\alpha_{nm}$ and $\beta_{nm}$. It is therefore the
main result regarding the two channel case.

Some special cases of (\ref{l_c:N=2}) may be of interest. For
isotropic backward scattering
$|\beta_{00}|^2=|\beta_{11}|^2 = 2 |\beta_{10}|^2$
follows $A=0$ and $B=0$ in (\ref{DGL:P,N=2}) and
the distribution function is independent of the
forward scattering: $P(\zeta) = 1/2$. This case corresponds
to the invariant distribution of an ensemble of unitary
$2 \times 2$ matrices. If the two channels are equivalent,
so that $|\beta_{00}|^2=|\beta_{11}|^2$ and thus $B=0$,
the distribution function $P(\zeta) \sim (1+A \zeta^2)^{-1}$
follows from (\ref{DGL:P,N=2}). The special case where
$|\beta_{00}|^2 = |\beta_{11}|^2 = |\beta_{10}|^2$
was already discussed in \cite{Dorokhov83} and is correctly
reproduced by our general solution.

\begin{figure}
\centerline{
\epsfxsize=0.47\textwidth
\epsfbox{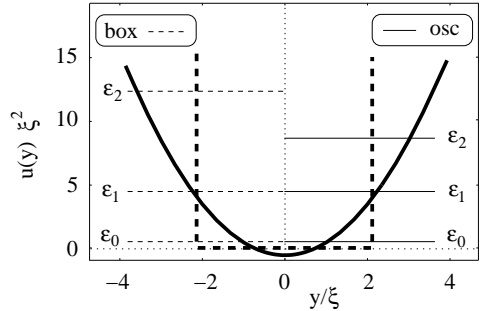}}
\caption{Two channel case $N=2$: The two
interaction potentials under consideration are
$u_{osc}(y)$ (thick solid line) and $u_{box}(y)$
(thick dashed line). The width of the box
and the absolut position of the oscillator
potential have been choosen so that the
energy levels
$\epsilon_{0,1}$ (thin lines) 
of the first two symmetric
bound states coincide.}
\label{fig:u_bereich}       
\end{figure}

We complete our discussion of the $N=2$ case with the
presentation of some numerical results for two specific
examples of interaction potentials.
Using the result (\ref{l_c:N=2}) one can analyze the
influence of the interaction potential $u(y)$ on
the localization length $l_c$.
The requirement of exactly two open channels
restricts the energy range to $\epsilon_1 < E < \epsilon_2$
where $\epsilon_0, \epsilon_1, \dots , \epsilon_n$
are the energies corresponding
to the bound states with given parity.
In order to demonstrate how the localization
length depends on the structure of the pair,
we present a comparison between the harmonic oscillator
interaction $u_{osc}(y)$ and the interaction where
the two particles are bound together by a box-like
potential $u_{box}(y)$. While $u_{osc}(y)$ is
a smooth function of the relative coordinate $y$ the
second interaction potential $u_{box}(y)$ is constant
for $|y|$ being smaller than some fixed value $y_s$
and jumps to infinity outside this region,
see Fig. \ref{fig:u_bereich}.

We first discuss the
case of bound states with even parity.
The size of the box and a
total energy shift for the harmonic oscillator
potential were adjusted so that the energy
levels $\epsilon_{0,1}$
entering (\ref{l_c:N=2}) are at the same
position. This approach ensures that all
direct energy dependencies of the localization
length have the same structure for either interaction
type. The energy window for the
$N=2$ case is then given by $4.5 < E \xi^2 < 8.5$ 
($4.5 < E \xi^2 < 12.5$) for the
harmonic oscillator (box-like
potential), Fig. \ref{fig:u_bereich}. 
%
%

\begin{figure}
\centerline{
\epsfxsize=0.47\textwidth
\epsfbox{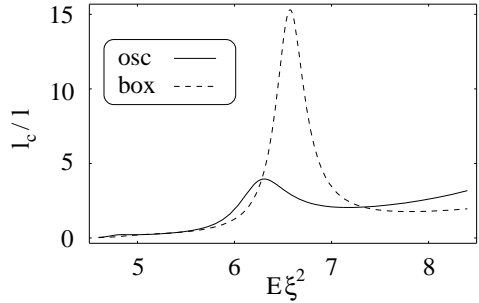}}
\caption{Two channel case $N=2$: The pair localization
length $l_c$ compared to the single particle localization
length $l$ as a function of the dimensionless
energy $E \xi^2$ (even parity of the bound state 
wave functions $\phi_n$).}
\label{fig:l_c_symm}
\end{figure}

The next step is to evaluate the matrix elements $\alpha$
and $\beta$ as given by (\ref{W_nm:alpha,beta}) and
(\ref{W_nm}). Inserting these matrix elements in
(\ref{l_c:N=2}) gives the result for
the localization length $l_c(E)$ which is presented
in Fig. \ref{fig:l_c_symm}. The interaction between the
constituents of the composed particle clearly
causes a delocalization compared to the
single particle with no internal degrees of
freedom. The reason is that the finite size of the
pair effectively smoothes out the disorder. However, 
this delocalization effect sensitively depends
on the total energy $E$ and the type of the two
particle interaction. The resonance in the
case of the box-like potential is clearly more
pronounced. If the energy is close to this resonance,
i.e. $E\xi^2 \approx 6.6$, the delocalization effect
differs by a factor $\sim 5$ due to the different
interaction potentials.
%
%

\begin{figure}
\centerline{
\epsfxsize=0.47\textwidth
\epsfbox{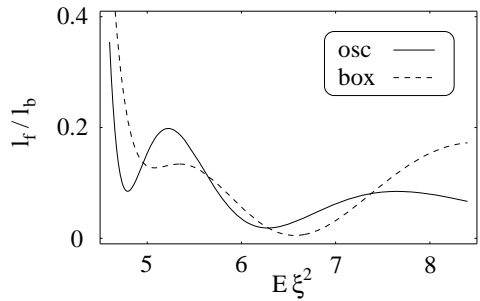}}
\caption{Two channel case $N=2$: The ratio between
forward scattering length $l_f$ and backward scattering length
$l_b$ as a function of the dimensionless
energy $E \xi^2$ (even parity of the wave functions
$\phi_n$).}
\label{fig:l_bf}
\end{figure}

Furthermore, we have examined the ratio between
forward and backward scattering $l_f / l_b$ as a function
of the energy. As Fig. \ref{fig:l_bf} indicates,
the forward scattering relative to the backward scattering
is already for just two channels enhanced. Although
the ratio $l_f/l_b$ depends on the energy this
effect holds true for the entire energy window
accessible for the two channel case. The suppression
of the backward scattering is because
the finite size of the pair reduces scattering processes with
large momentum transfer as can be concluded from
Eqs. (\ref{W_nm:alpha,beta}, \ref{W_nm}).
This effect is even more pronounced if many
channels are open, a fact that is of importance
for the solution of the multi-channel case, as discussed in
Sec. \ref{Multi-channel}.
%
%

\begin{figure}
\centerline{
\epsfxsize=0.47\textwidth
\epsfbox{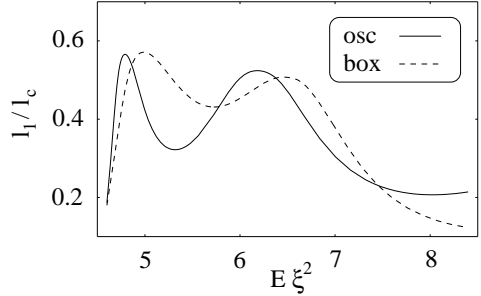}}
\caption{Two channel case $N=2$: The ration
$l_1 / l_c$ of the length scales of the two channels
as a function of the dimensionless
energy $E \xi^2$ (even parity of the wave functions
$\phi_n$).}
\label{fig:l_c1_symm}
\end{figure}

Using the sum rule (\ref{avg.l_n}) we calculate the
length scale $l_1$ of the transmission
eigenvalue of the second internal channel.
It is given by $l_1 = (2 l_b^{-1} - l_c^{-1})^{-1}$,
see Fig. \ref{fig:l_c1_symm}. 
We find that the ratio $l_1/l_c$ is smaller
than $0.6$ over the entire energy range for either
interaction potential. This justifies
the assumption that the transmission is mainly due to
one internal channel if the system is large,
i.e. $L \gg l$, because of the exponential
dependence of the transmission amplitudes
on the system size.
%
%

\begin{figure}
\centerline{
\epsfxsize=0.47\textwidth
\epsfbox{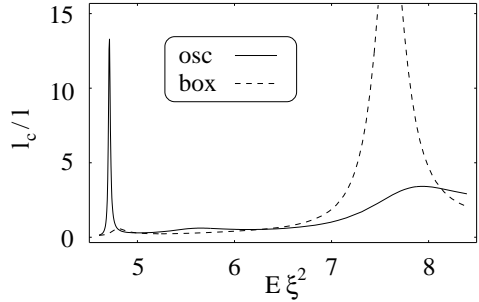}}
\caption{Two channel case $N=2$: The pair localization
length $l_c$ compared to the single particle localization
length $l$ as a function of the dimensionless
energy $E \xi^2$ (odd parity of the wave functions
$\phi_n$).}
\label{fig:l_c_anti}
\end{figure}

Finally we compute the localization length for the
situation where the two bound states are of odd parity.
Again, the parameters of the interaction potentials
are fixed such that the energies of the two involved
energy levels coincide for the different
interaction potentials. The resulting localization
length is plotted in Fig. \ref{fig:l_c_anti}. There
is a significant difference between
the localization lengths in the case of wave functions with
even parity compared to odd parity.
For the harmonic oscillator potential
there is a sharp resonance near the lower limit
$\epsilon_1$ of the energy window in the second case
(Fig. \ref{fig:l_c_anti}) while the first case shows
a smooth behavior with a weak resonance in the middle
of the energy range (Fig. \ref{fig:l_c_symm}).
For the $u_{box}(y)$ interaction the resonance
position and height also changes significantly.
Since the energy levels are
shifted to the same position for either case the
differences result only from the different structure of the
bound state wave functions. The largest
delocalization effect occurs for $u_{box}(y)$
for the states of odd parity and is about $l_c/l \approx 25$
at the resonance.
%
%

\section{Conclusion}
\label{Conclusion}
In this work we have studied the localization
of a pair of two bound particles in a one-dimensional
system with weak disorder. The interaction potential
within this composite particle
is fully taken into account within the
semiclassical approximation. Our results are based
on a method introduced by Dorokhov for two particles
bound by a harmonic oscillator potential.
We have generalized the method from this special interaction
potential to an arbitrary attractive interaction.
This allows us to analyze the dependence of the
pair localization on the interaction. We find an 
enhancement of the pair localization length $l_c$
in comparison to the single particle mean free path $l$.
This enhancement is independent of the form of the
pair interaction
potential in the limit of many bound states $N \gg 1$. It is given
by $l_c/l \sim N$ as can be seen from the central result
(\ref{final_result:l_c}) in conjunction with (\ref{WKB:l_b,result}).
Furthermore we derived an exact solution
for the two channel case $N=2$. For a bound pair with 
$N=1,2$ we observe a 
sensitive dependence of the localization length 
(\ref{l_c:N=1}), (\ref{l_c:N=2}) on 
the shape of the interaction potential,
the parity of the involved bound states and the kinetic energy
of the pair. We expect that such a behavior is typical
for bound pair with a small number of bound states.

The approach described in this work is valid for
two identical particles. Nevertheless, it is possible to
extend the method without difficulties to a pair
of different particles (e.g. electron-hole pair). In this case,
the expression for the matrix elements (\ref{W_nm})
has to be modified taking into account that the random
potential and the mass may be different for the two particles.
However, the described method cannot be extended to
a repulsing interaction because the assumption
that the pair size is smaller than the mean free path
breaks down.


\end{document}